\documentclass[showpacs]{revtex4}

\usepackage{amssymb}
\usepackage{amsfonts}
\usepackage{amsmath}
\usepackage{epsfig}
\usepackage{xspace}
\usepackage{graphicx}
\usepackage[latin1]{inputenc}

\def\ro{\mbox{\boldmath $\rho$}}
\def\beq{\begin{eqnarray}}
\def\eeq{\end{eqnarray}}
\def\f{\frac}

\def\bit{\begin{itemize}}
\def\eit{\end{itemize}}

\def\cal{\mathcal}

\begin{document}
\title{A discrete finite-dimensional phase space approach for the description of Fe8 magnetic clusters: Wigner and Husimi functions}
\author{Evandro C. Silva and Di\'{o}genes Galetti}
\affiliation{Instituto de F\'{\i}sica Te\'{o}rica - S\~{a}o Paulo State University\\
						 Rua Pamplona 145, 01405-900, S\~{a}o Paulo, SP, Brazil \\
             E-mail address: mandrake@ift.unesp.br, galetti@ift.unesp.br}
\date{\today}

\begin{abstract}
A discrete quantum phase space formalism is used to discuss some basic aspects
of the spin tunneling occurring in Fe8 magnetic clusters by means of  Wigner
functions as well as Husimi distributions. Those functions were obtained for
sharp angle states and symmetric combinations of the lowest energy doublet -- the particular tunneling energy doublet --
with the application of an external magnetic field. The time
evolution of those functions carried out numerically allows one to extract
valuable information about the dynamics of the states under consideration,
in particular, the coherent oscillations associated with the particular energy doublet. It also shown that
an entropy functional can be constructed out of the discrete Husimi distribution and that the analysis of those
coherent oscilations process allows one to clarify the meaning of the entropy in this case.
\end{abstract}

\maketitle
\section{Introduction}
The basic ideas governing the discrete phase space picture of quantum
mechanics of physical systems described by finite-dimensional state spaces
have been long established, and the concept of discrete Wigner functions and
the discrete analogue of the Weyl-Wigner transformations are now recognized
as important tools in dealing with those systems \cite{woo, gapi,cohe}.
Within this description, the quasiprobability distribution functions are
writen in terms of a pair of discrete variables -- for each degree of freedom --
whose domains are finite sets of integer numbers which constitute a finite
lattice that stands for a quantum discrete phase space. In this context,
some attempts were also made in order to present a unified approach to the
discrete quasiprobability distribution functions in the sense of obtaining
the Glauber-Sudarshan, Wigner and Husimi functions as particular cases of
s-parametrized discrete phase space functions \cite{opa,maumarga,marmauga}.
In particular it was then shown that a hierarchical order among them through a
smoothing process characterized by a discrete phase space function that
closely resembles the role of the Gaussian function in the continuous phase
space can be established \cite{marmauga}. From an operational point of view,
it is clear that if we have a discrete Wigner function, the corresponding discrete Husimi
distribution can also be directly obtained. Furthermore, discrete quasiprobability 
distribution functions in finite phase spaces have
potential applications in quantum
state tomography \cite{marmauga,leon,miquel}, quantum teleportation 
\cite{marmauga,bennet,koniorkzyk,paz,ban}, phase space representation of quantum 
computers \cite{miquel2}, open quantum systems \cite{bianucci}, quantum information
theory \cite{paz2}, and quantum computation \cite{galvao}, where they are a
natural tool to deal with the essential features of the inherent kinematics
of the physical systems.

In the present paper we intend to show that the discrete phase space
approach presented before \cite{gapi,maumarga,marmauga,gapi2} can also be applied to
the case of magnetic molecules, in particular to the Fe8 magnetic cluster \cite{wieg},
where certain important quantum phenomena, such as spin tunneling, are currently 
considered because they are based on 
strong evidences coming from results of measurements with great resolution. 
That magnetic cluster has been deeply
investigated in the last years \cite{barra,sangregorio,caneschi} and has been
characterized as an useful tool in studying the importance of quantum
effects in spin tunneling. Among the interesting properties it presents, it
must be remarked that it has a $J=10$ spin associated with its magnetic
moment, and that the anisotropy constants characterizing its structure are
experimentally determined which allows one to construct a \textit{bona fide}
phenomenological Hamiltonian from which spectra and wave functions can be
obtained. Besides, since it presents a well determined cross-over
temperature below which the tunneling of the magnetic moment is dominated by
the quantum effects \cite{sangregorio}, the experimental data obtained in
that temperature domain can give inestimable information about that quantum
phenomenon. The application of external magnetic fields on the Fe8 cluster
also gives additional information about its quantum behavior and reveals
interesting features of the energy spectra \cite{werns}. In this form, the
quantum kinematical content as well as the proposed phenomenological
Hamiltonian associated with this magnetic cluster provide the basic elements
from which we can develop and apply the discrete phase space approach. 
As a
consequence, we show that Wigner functions, as well as Husimi distributions,
can be obtained that allow us to describe the behavior of the tunneling energy doublet
 and to study the effects of the application of external
magnetic fields. Furthermore,  after a Wehrl's type entropy functional has been constructed
out of the Husimi distribution, a clear interpretation of that entropy is achieved through the
study of the behavior of the tunneling energy doublet. It is shown
that the discrete phase space approach results are in good agreement with
the direct numerical calculations.

This paper is organized as follows. In Section II we present a brief review
of the discrete phase space approach, including the procedures through which
we obtain the discrete Wigner functions as well as the Husimi distributions.
The procedure of getting the time evolution of the Wigner function is also
presented. The application of the formalism to the Fe8 magnetic cluster is
carried out in Section III, and Section IV is devoted to the conclusions.
\section{Discrete phase space formalism: a brief review}
The possibility of describing nonclassical states and expectation values of quantum systems
through a quantum phase space picture has been widely explored in the
literature mainly in what concerns those physical systems described by
continuous variables \cite{todos}. In this section we want to recall the main
aspects of a possible quantum phase space picture especifically constructed
for treating physical systems described by finite-dimensional state space.
\subsection{Wigner functions and the Liouvillian}
In the past, we have proposed a mapping scheme that allows to represent
any operator acting on a finite-dimensional state space as a function of
integer arguments \cite{gapi,gapi2} on a discrete $N^{2}$-dimensional phase
space; in other words, we have extended the well-known Weyl-Wigner procedure \cite{todos}
of operators mapping  for finite $N$-dimensional Hilbert state
space cases. In this form, we are then able to study the quantum properties
of a given system of interest through the mapping of the relevant
operators and studying the corresponding mapped functions. Of course we can
also obtain discrete Wigner functions that represent state operators upon a
discrete $N^{2}$-dimensional phase space \cite{gapi}. In this connection, it
is direct to see that we can also construct the Husimi distribution
function, as well as the analogue of the Glauber-Sudarshan function if this
is desired \cite{maumarga,marmauga}.

From a general point of view, let us start by considering that, for the systems under 
consideration, the discrete Weyl-Wigner transformation of a given operator $\mathbf{O}
$ into its corresponding representative in the discrete quantum
phase space is given by a trace operation \cite{gapi,gapi2}
\begin{equation*}
O\left( m,n\right) =\frac{1}{N}Tr\left[ \mathbf{G}^{\dag }\left( m,n\right) 
\mathbf{O}\right] .
\end{equation*}
In this expression $\mathbf{G}\left( m,n\right) $ is an operator basis which
guarantees the mod $N$ invariance in the discrete phase space, and is given by
\begin{equation}
\mathbf{G}\left( m,n\right) =\sum_{k,l=-\ell}^{\ell}\frac{\mathbf{S}\left(
k,l\right) }{\sqrt{N}} e^{\left[ i\pi \phi \left( k,l;N\right) -\frac{2\pi i
}{N}\left( mk+nl\right) \right]} ,  \label{oper-base}
\end{equation}
where $\mathbf{S}\left( k,l\right) $ is the symmetrized, orthonormal and
complete operator basis introduced by Schwinger \cite{schwinger} and 
$-(N-1)/2\leq \ell\leq (N-1)/2$.

It is also well known that the time evolution of a quantum system can be
described by the von Neumann-Liouville equation for the density operator.
Now, with the help of the expression for the mapped commutator of two
operators, it is straightforward to obtain the mapping of that dynamical
equation on the corresponding discrete quantum phase space equation. As
already shown before \cite{gapi,gapi2}, the von Neumann-Liouville time
evolution equation (we will consider hereafter $\hbar =1$)
\begin{equation*}
i\frac{\partial }{\partial t}\ro=\left[ \mathbf{H},\ro\right]
\end{equation*}
is mapped onto
\begin{equation}
i\frac{\partial }{\partial t}\cal{W}\left( u,v;t\right) =\sum_{r,s=-\ell}^{\ell}
\mathcal{L}\left( u,v,r,s\right) \cal{W}\left( r,s;t\right) ,  \label{vNL}
\end{equation}
where
\beq
\nonumber \cal{L}(u,v,r,s) &=& 2i\sum_{m,n,a,b,c,d=-\ell}^{\ell}\f{h(m,n)}{N^4} 
\\\nonumber &\times& \sin\left[\f{\pi}{N}(bc-ad)\right] e^{[i\pi\Phi(a,b,c,d;N)]}
\\ & \times& e^{\left\{\f{2\pi i}{N}[a(u-m)+b(v-n)+c(u-r)+d(v-s)]\right\}}\label{liou}\eeq
is identified with the discrete mapped expression of the Liouvillian of the
system, $h\left( m,n\right) $ stands for the mapped expression of the here
assumed time-independent Hamiltonian of interest,
and the phase $\Phi(a,b,c,d;N)$ guarantees the mod N invariance. The function of integers 
$\cal{W}\left( r,s;t\right) $ is the discrete Wigner function which is
obtained by mapping the density operator defined in the finite-dimensional
state space; for instance, for a pure state
\begin{equation*}
\ro\left( t\right) =\mid \psi \left( t\right) \rangle \langle
\psi \left( t\right) \mid
\end{equation*}
we have
\begin{equation*}
\cal{W}\left( m,n;t\right) =\frac{1}{N}Tr\left[ \mathbf{G}^{\dag }\left(
m,n\right) \mid \psi \left( t\right) \rangle \langle \psi \left( t\right)
\mid \right].
\end{equation*}

With these assumptions and considering the time evolved density operator to
be written in the form of a series \cite{garu}
\beq
\nonumber \ro\left( t\right) &=& \ro\left( t_{0}\right) -i\left(t-t_{0}\right)
\left[ \mathbf{H},\ro\left( t_{0}\right) \right]
\\ &+& i^{2}\frac{\left( t-t_{0}\right) ^{2}}{2!}\left[ \mathbf{H},\left[ \mathbf{H
},\ro\left( t_{0}\right) \right] \right] -\ldots ,  \label{serie}
\eeq 
we see that its mapped expression is given by
\beq  \nonumber
\cal{W}\left( u,v;t\right) &=&\sum_{r,s=-\ell}^{\ell}\cal{W}\left( r,s;t_{0}\right)\Big\{ \delta _{r,u}^{\left[ N
\right] }\delta _{s,v}^{\left[ N\right] }
\\\nonumber
&+& (-i)\left(t-t_{0}\right) \mathcal{L}\left( u,v,r,s\right) + \frac{\left( -i\right)^{2}\left( t-t_{0}\right) ^{2}}{2!}
\\ 
&\times& \sum_{x,y}\mathcal{
L}\left( u,v,x,y\right) \mathcal{L}\left( x,y,r,s\right) +...\Big\}, \label{serie2} 
\eeq
where now $\cal{W}\left( u,v;t\right) $ is the solution at time $t$ of the
von Neumann-Liouville equation for the Wigner function. In expression (\ref%
{serie2}) those terms between curly brackets can be identified as the mapped
propagator that associates the initial Wigner function at time $t_{0}$ with
a time evolved Wigner function at time $t$. We also see that the time
evolution expression is constituted of sums of products of the array
associated with the Liouvillian and the discrete Wigner functions taken over
the discrete phase space. Then, from a numerical point of view, to calculate
the time evolution of a particular density operator we only have to
construct the arrays associated with the Liouvillian and the Wigner function
of interest at time $t_{0}$.
\subsection{The Husimi distribution function}
From another perspective, in what concerns the mapped expression of the
density operator, we can also have at our disposal a discrete Husimi
function. The formal way of obtaining this function in the present context
was discussed in previous papers in which the continuous Cahil-Glauber
formalism \cite{cahil-glauber} was extended for finite-dimensional state
spaces \cite{maumarga,marmauga}. Let us briefly review how to obtain the
discrete Husimi distribution from a previously given discrete Wigner function.
This can be accomplished by means of an s-parameterized mod$(N)$
operator basis, which is given as
\beq
\nonumber\mathbf{T}^{\left( s\right) }\left( m,n\right) =\sum_{\eta
,\xi=-\ell }^{\ell} e^{\left[ i\pi \phi \left( \eta ,\xi ;N\right) -\frac{2\pi i}{N}
\left( \eta m+\xi n\right) \right]} \frac{\mathbf{S}^{\left( s\right) }\left( \eta,\xi \right)}{\sqrt{N}} .  \label{kernel-t}
\eeq
Here the new mod$(N)$ invariant operator basis, $T^{\left( s\right)}\left( m,n \right)$,
is defined through a discrete Fourier transform of the extended operator basis, namely
\begin{equation}
\mathbf{S}^{\left( s\right) }\left( \eta ,\xi \right) =\left[ K\left( \eta
,\xi \right) \right] ^{-s}\mathbf{S}\left( \eta ,\xi \right) ,
\label{base-s}
\end{equation}
where $K\left( \eta ,\xi \right) $ is a bell shaped function expressed as a
product of Jacobi theta functions
\beq
\nonumber K(\eta,\xi)&=&\left\{2\left[\vartheta_3(0|ia)\vartheta_3(0|4ia) \right. \right.
\\ \nonumber &+& \left. \left. \vartheta_4(0|ia)\vartheta_2(0|4ia)\right]\right\}^{-1}\Big\{\vartheta_3(\pi a\eta|ia)\vartheta_3(\pi a \xi|ia)  
\\ \nonumber &+&  \vartheta_3(\pi a \eta|ia)\vartheta_4(\pi a \xi|ia)e^{(i\pi\eta)} 
\\  \nonumber &+&  \vartheta_4(\pi a \eta|ia)\vartheta_3(\pi a \xi|ia)e^{(i\pi\xi)}   
\\ &+&  \nonumber \vartheta_4(\pi a \eta|ia)\vartheta_4(\pi a \xi|ia)e^{[i\pi(\eta+\xi+N)]}\Big\}
\eeq
with $a=\left( 2N\right) ^{-1}$ and $\mathbf{S}\left( \eta ,\xi \right)$ are the Schwinger basis elements already mentioned before. The complex parameter $s$ that appears in
the above expression is limited to the interval $|s|\leq 1$.

In analogy with the decomposition expression of an operator into an operator basis, what involves its
representative defined on the discrete quantum phase space, we have now%
\begin{equation*}
\mathbf{O}=\frac{1}{N}\sum_{m,n=-\ell}^{\ell}O^{(-s)}\left( m,n\right) \mathbf{T}^{\left(
s\right) }\left( m,n\right) .
\end{equation*}%
In particular, if we choose $\mathbf{O}=\ro$\textbf{\ } we see that for $s=0$
this expression gives the mapping leading to the Wigner function, while for
$s=-1$ it leads to the Husimi distribution.

Within this formalism we also have the expression that allows us to obtain the
Husimi distribution directly from the Wigner function, namely%
\begin{equation*}
\cal{H}\left( m,n\right) =\sum_{m^{\prime },n^{\prime }=-\ell}^{\ell}E\left( m^{\prime
}-m,n^{\prime }-n\right) \cal{W}\left( m^{\prime },n^{\prime }\right) ,
\end{equation*}%
where $E\left( m^{\prime }-m,n^{\prime }-n\right) $ characterizes the
mapping kernel of the transformation, and it is given by 
\begin{equation*}
E\left( m^{\prime }-m,n^{\prime }-n\right) =Tr\left[ \mathbf{T}^{\left(
0\right) }\left( m,n\right) \mathbf{T}^{\left( -1\right) }\left( m^{\prime
},n^{\prime }\right) \right] .
\end{equation*}

In what concerns the time evolution, it is then clear that, once the Wigner
function is given at any instant of time, the Husimi distribution can be
obtained straightforwardly in the same way. Furthermore, in what relates to
the system state, since the Husimi distribution has a well-behaved
shape, we expect to obtain a clearer picture of the physical process of interest
through its use.
\section{Aplication: Fe8 magnetic cluster}
The discovery of magnetic molecules, in particular the Fe8 magnetic cluster,
which was first synthesized by Wieghardt et al \cite{wieg}, was a great
contribution for the study of quantum effects in the dynamics of spin
tunneling \cite{monte}. Under proper temperature conditions, the pure quantum
effects in the dynamics of the magnetization become manifest in some
measurements performed with this molecule, specially due to the very well
measured constants characterizing its structural properties. The study of
the orientation of the magnetic moment with these clusters also benefits from
the fact that the presence of an external magnetic field may enhance some
quantum transitions that allow one to test basic hypothesis about the
quantum spin tunneling.

The main experimentally determined results related to the Fe8 magnetic
cluster indicate that it is a system with a $J=10$ spin -- that implies in a 
$21$-dimensional state space for this degree of freedom-- with an observed
barrier of about $24\ K$. Furthermore, it was verified that below $0.35\ K$
the relaxation of the magnetization becomes temperature independent \cite%
{sangregorio}, and that when an external magnetic field with intensity that
is given by integer multiples of $\Delta H_{\parallel }\simeq 0.22\ T$ is
applied along the easy-axis of the cluster (z-axis) the hysteresis curve has
a behavior \cite{caneschi} which can be interpreted as if there is an energy
matching of states on both sides of a double-minima barrier, similar to what
had already been proposed in another magnetic molecule, namely the Mn12ac
cluster \cite{fried}. From the experimental data the proposed Hamiltonian for
the Fe8 cluster (sometimes called giant spin model) in the presence of
external magnetic fields has the form%
\beq
\label{hamiltoniana}
\mathcal{\textbf{H}} = {\textbf{H}}_0 + {\textbf{H}}_1  ,
\eeq
where the term 
\begin{equation*}
\mathcal{\textbf{H}}_{0}\mathcal{=}D\mathbf{J}_{z}^{2}+\frac{E}{2}\left( \mathbf{J}%
_{+}^{2}+\mathbf{J}_{-}^{2}\right) 
\end{equation*}
constitutes the part that is established by the measured properties of the molecular
structure, and the remaining ones 
\begin{equation*}
\mathcal{\textbf{H}}_{1}\mathcal{=}A\mathbf{J}_{z}+B\left( \mathbf{J}_{+}+\mathbf{J}%
_{-}\right) +C\left( \mathbf{J}_{+}-\mathbf{J}_{-}\right) 
\end{equation*}
give the contribution that is associated with the external magnetic field.
The anisotropy constants are \cite{caneschi} $D/k_{B}=-0.275\ K$ and $%
E/k_{B}=0.046\ K$, where $k_{B}$ is the Boltzmann constant, while $A=g\mu
_{B}H_{\parallel }$, $B=g\mu _{B}\cos \alpha H_{\perp }$, and $C=g\mu
_{B}\sin \alpha H_{\perp }$ are the parameters associated with the magnetic
field intensity along the z-axis (easy-axis), y-axis (medium axis) and
x-axis (hard-axis) respectively. Furthermore, this model Hamiltonian is expressed in terms of the angular momentum
operators that obey the standard commutation relations%
\begin{equation*}
\left[ \mathbf{J}_{\pm },\mathbf{J}_{z}\right] =\mp \mathbf{J}_{\pm },
\qquad \left[ \mathbf{J}_{+},\mathbf{J}_{-}\right] =2\mathbf{J}_{z},
\end{equation*}
and its numerical diagonalization can be directly carried out within the $21$%
-dimensional set of $|j,m\rangle $ states, so that those energy eigenvalues
thus obtained can be useful as reference values.

Now, the main requirement for the use of the proposed description of physical systems with
finite-dimensional state space  consists in characterizing
the starting space state such that the quantum phase space mapping scheme is completely established.
In the present case we have to establish a connection between our approach
and the mathematical description of the proposed model for the Fe8 magnetic
cluster. This connection, which is then the key element in order to allow
the use of the quantum phase space description for the Fe8 cluster, is
obtained just by choosing%
\begin{equation*}
|u_{k}\rangle \equiv |j,k\rangle
\end{equation*}%
with the associated eigenvalue equation following the prescription proposed by Schwinger \cite{schwinger}
\begin{equation}
\mathbf{U}\ |u_k\rangle =\exp \left( 2\pi i\frac{k}{N}\right) |j,k\rangle ,
\label{autoU}
\end{equation}%
where $\mathbf{U}$ is an unitary operator and $|u_{k}\rangle$ its eigenstates. 
Furthermore,
\begin{eqnarray*}
\mathbf{J}_{z}|j,k\rangle &=&k|j,k\rangle , 
\\ \mathbf{J}^{2}|j,k\rangle &=&j(j+1)|j,k\rangle ,
\\ \mathbf{J}_{\pm }|j,k\rangle &=&\sqrt{\left( j\mp k\right) \left( j\pm k+1\right) }|j,k\pm 1\rangle ,
\end{eqnarray*}
as usual, where $-10\leq k\leq 10$.

Therefore, we have a $21$-dimensional space of the $\mathbf{U}$ eigenvectors as given
above, and consequently we also have another $21$-dimensional space, $\{\left|v_l\right\rangle\}$, of
eigenvectors of $\mathbf{V}$, the complementary unitary operator also proposed by Schwinger.
It is worth noting that the set of states $\left\{|v_{l}\rangle \right\} $, 
where $-10\leq l\leq 10$, is in fact a basis in this finite space since it is
orthonormal and complete, thus it is also a \textit{bona fide} basis. 
Since it was also shown \cite{schwinger} that the set of states $\left\{|v_{l}\rangle \right\}$ is obtained through
a discrete Fourier transform of the basis states $\left\{|u_{k}\rangle \right\}$, 
the physical interpretation of the $\left\{|v_{l}\rangle \right\}$ states is then direct: they are
associated with the polar angle orientation of Fe8 cluster spin.

Based on this association, we can now use the discrete Weyl-Wigner
formalism discussed before to describe the quantum behavior of the Fe8
cluster in a discrete $21^2$-dimensional phase space. Thus far the
kinematical content of the description has been already set up, then we can
go one step further and obtain the mapped expression associated with the Fe8
cluster Hamiltonian (\ref{hamiltoniana}). Following the mapping scheme
presented before the discrete phase space expression for that Hamiltonian is 
\beq
\nonumber  h\left( m,n\right)&=&Am+Dm^{2}+EQ\left( m,n\right) \cos \frac{4\pi n}{N}
\\
\nonumber &+&\sum_{k,r=-10}^{10}\f{e^{\f{2\pi i}{N}\left[ r\left( m-k-\frac{1}{2}\right) -n\right]}}{2N}
\\
\nonumber &\times& \sqrt{\left( j-k\right) \left( j+k+1\right) } \left( B\cos \alpha +C\sin\alpha \right) 
\\
\nonumber &+& \sum_{k,r=-10}^{10}\f{e^{\f{2\pi i}{N}\left[ r\left( m-k+\frac{1}{2}\right) -n\right]}}{2N}
\\
\nonumber &\times& \sqrt{\left( j+k\right) \left( j-k+1\right) }\left( B\cos \alpha -C\sin
\alpha \right) ,
\eeq
where 
\begin{equation*}
Q\left( m,n\right) =\sqrt{\left( j+m\right) \left( j-m+1\right) \left(
j-m\right) \left( j+m+1\right) }.
\end{equation*}%
With this result, it is straightforward now to find the expression for the
Liouvillian. After a direct but tedious calculation, the discrete phase
space expression for the Liouvillian is obtained, and it is written as 
{\small\beq
\nonumber \mathcal{L}(u,v,r,s)&=&\f{2i}{N^3}\sum_{m,a,c,d}\left(\left[-(Am+Bm^2)\sin{\left(\f{\pi}{N}ad\right)} \right. \right.
\\\nonumber &\times& \left. \left. e^{i\pi\Phi(a,0,c,d;N)} \right. \right.
\\\nonumber &+& \left. \left.  C\sqrt{\left(j+m\right)\left(j-m+1\right)}\sin{\left[\f{\pi}{N}(c-ad)\right]} \right.\right.
\\\nonumber &\times& \left. \left. e^{\f{i\pi}{N}(a+2v)}e^{\f{2i\pi}{N}c(r-a)} e^{i\pi\Phi(a,1,c,d;N)}\right.\right.
\\\nonumber &-& \left. \left. D\sqrt{\left(j-m\right)\left(j+m+1\right)}\sin{\left[\f{\pi}{N}(c+ad)\right]}\right.\right.
\\\nonumber &\times& \left. \left. e^{\f{-i\pi}{N}(a+2v)}e^{\f{2i\pi}{N}c(r-a)} e^{i\pi\Phi(a,-1,c,d;N)} \right.\right.
\\\nonumber &+& \left. \left.  \f{Q(m,n)}{2} \right.\right.
\\\nonumber &\times& \left. \left. \left\{\sin{\left[\f{\pi}{N}(2c-ad)\right]}e^{\f{4\pi i}{N}v}e^{i\pi\Phi(a,2,c,d;N)}\right. \right. \right. 
\\\nonumber &-& \left.\left.\left. \sin{\left[\f{\pi}{N}(2c+ad)\right]}e^{\f{-4\pi i}{N}v}e^{i\pi\Phi(a,-2,c,d;N)}\right\} \right]\right.
\\ &\times& \left. e^{\f{2\pi i}{N}[a(u-m)+c(u-r)+d(v-s)]}\right). 
\label{liouville}
\eeq}

Having obtained the discrete Wigner function and using the mapped
series, Eq. (\ref{serie2}), we are then able to compute the temporal evolution of a
state given at $\tau=0$ with the expression above, Eq. (\ref{liouville}).

In what concerns the Wigner function at $\tau=0$, we may choose those that may
be relevant for our purposes. At first, we start, for simplicity, from the
density operator $\mathbf{\ro }$ for a pure state%
\begin{equation*}
\ro=|\psi \rangle \langle \psi |,
\end{equation*}%
where $|\psi \rangle $ can be expressed in terms of the eigenstates of the
operator $\mathbf{U}$, Eq. (\ref{autoU}), which are also, due to the adopted
formalism, eigenstates of the angular momentum operators $\mathbf{J}^{2}$ and $\mathbf{J}_{z}$%
, namely%
\begin{equation*}
|\psi \rangle =\sum_{l=-10}^{10}C_{l}|j,l\rangle .
\end{equation*}%
As has already been shown elsewhere \cite{gapi3}, the Wigner function in this
case is written as%
\begin{equation}
\cal{W}\left( m,n\right) =\frac{1}{N}\sum_{r,s=-10}^{10}\exp \left[ \frac{2\pi i}{N}%
\left( mr+ns\right) \right] g\left( r,s\right) ,  \label{wigner1}
\end{equation}%
where 
\begin{equation*}
g\left( r,s\right) =\frac{1}{N}\sum_{k=-10}^{10}C_{k}C_{\left\{ k+s\right\} }^{\ast
}\exp \left[ -\frac{2\pi i}{N}r\left( k+\frac{s}{2}\right) \right] ,
\end{equation*}%
and the symbol $\left\{ k+s\right\} $ stands for the sum cyclically
restricted to the interval of labels, e.g., $k_{\max }+1=k_{\min }$. As can
be directly recognized, a great advantage of using the Wigner functions in
this form is that, by diagonalizing the Hamiltonian in the angular momentum
state basis $\left\{ |j,k\rangle \right\} $, we can obtain the coefficients $%
C_{k}$ to construct the Wigner function through Eq. (\ref{wigner1}).
\subsection{Sharp angle state}
A first illustrative case that can be directly treated is a sharp angle
state associated with the physical configuration where, at $\tau=0$, the
magnetic moment of the Fe8 cluster points to a particular angle 
$\left\{ \theta ,\varphi \right\}$. The parallel and transverse 
magnetic fields are, in particular, $H_{\parallel}=0.01\ T$, and 
$H_{\perp }=2.0\ T$ respectively, being  $H_{\perp }$ oriented at an angle of 
$\varphi=\pi/4$ relative to the $x$-axis, and the initial state, a sharp angle,
is peaked at $n=0$, where $\theta_n=\f{\pi}{10}n$ with $-\pi\leq\theta_n\leq\pi$, and $-10\leq n,m\leq 10$. The correspondent sequence of the time evolved 
Wigner functions is depicted in Fig. 1. 
\begin{figure}[h!]

\includegraphics[width=5.85cm]{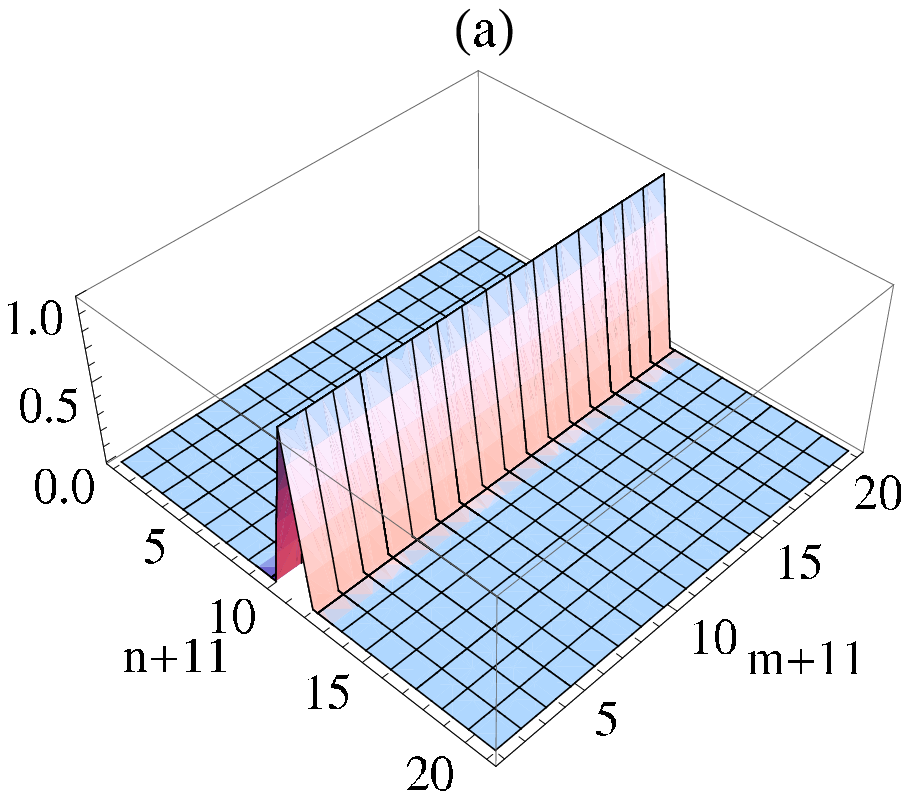}
\includegraphics[width=5.85cm]{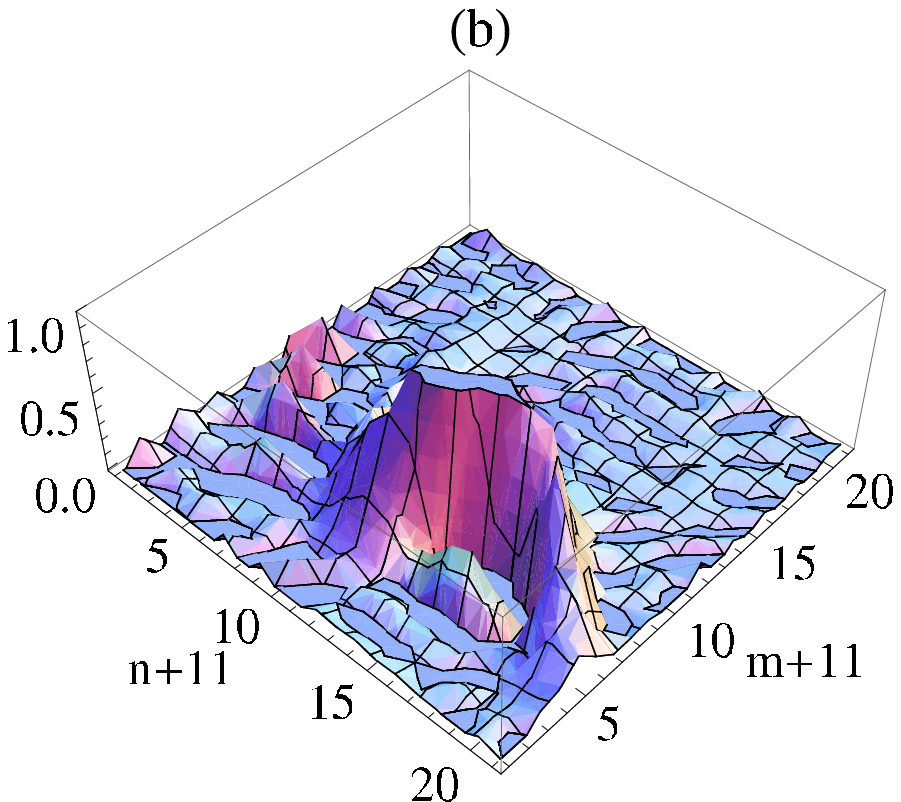}
\includegraphics[width=5.85cm]{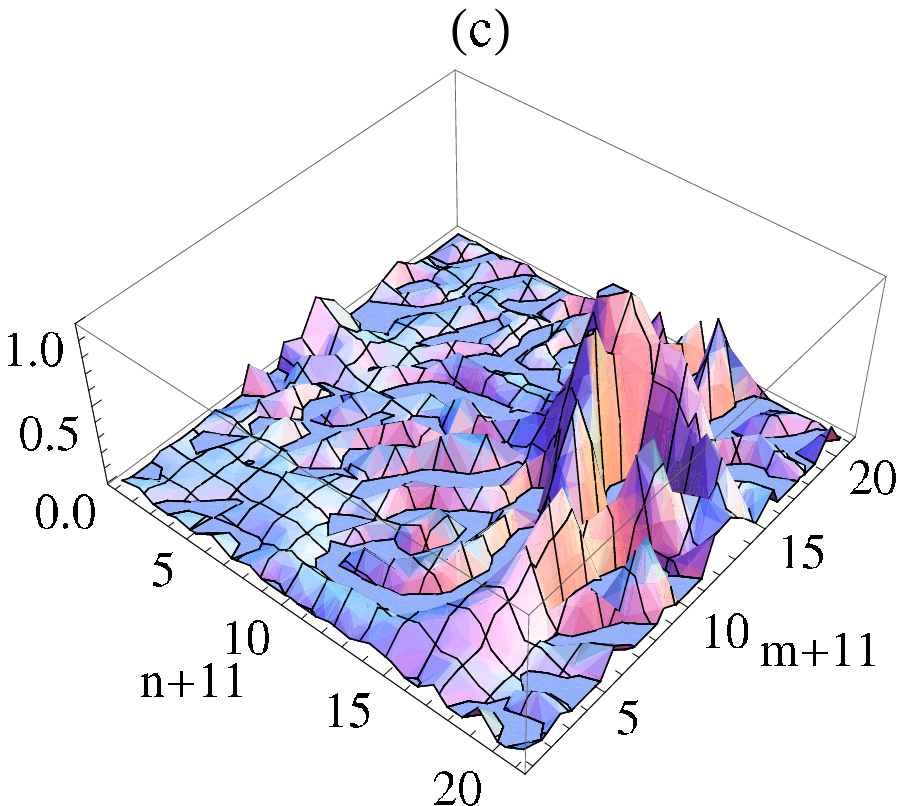}
\includegraphics[width=5.85cm]{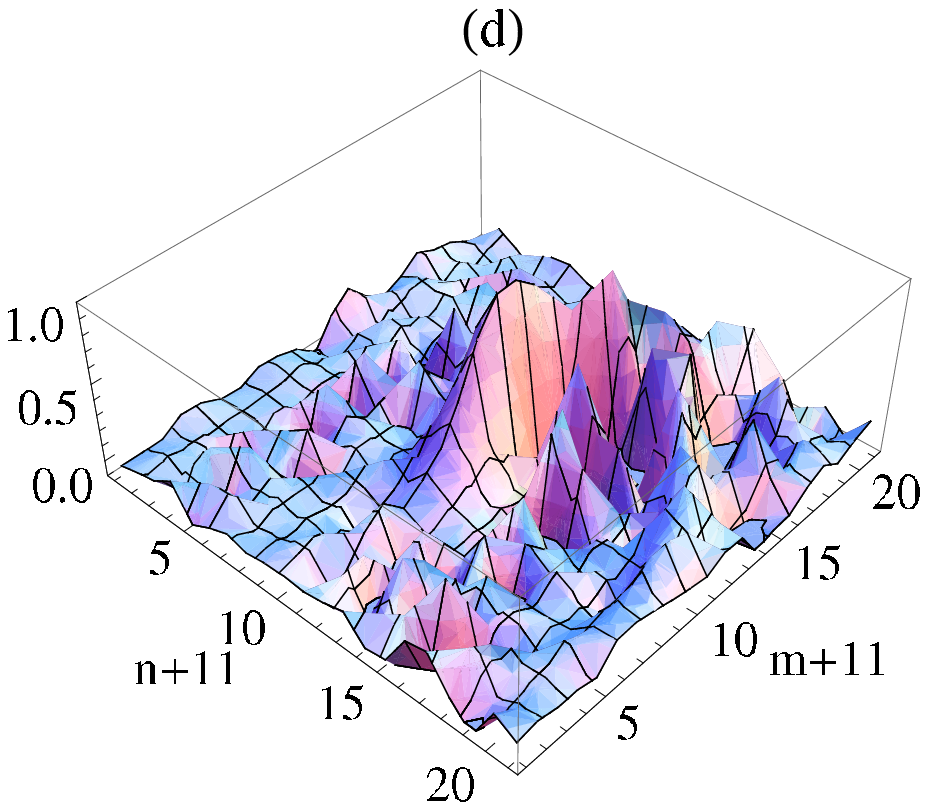}

\caption{Time evolved Wigner function calculated with time steps of 
$\Delta$t = $0.05 \times 2\pi$ $K^{-1}$ and external magnetic fields 
$H_{\parallel} = 0.01$ T and $H_{\perp} = 2.0$ T oriented at an angle 
of $\phi=\pi/4$ with the $x$-axis. The labels m and n correspond to the
finite-discrete angular momentum and angle variables respectively displaced 
11 units of their original intervals. Figure (a) corresponds to the Wigner
function at $\tau=0$. Figures (b), (c), and (d) depict the Wigner functions
for $\tau = 0.4\times 2\pi$ $ K^{-1}$, $0.8\times 2\pi$ $ K^{-1}$ and 
$1.2\times 2\pi$ $K^{-1}$ respectively.}

\end{figure}

Now, since we have the time evolved Wigner functions for the peaked angle state, the associated Husimi distributions can be directly obtained, and are shown in Fig. 2
\begin{figure}[h!]

\includegraphics[width=5.85cm]{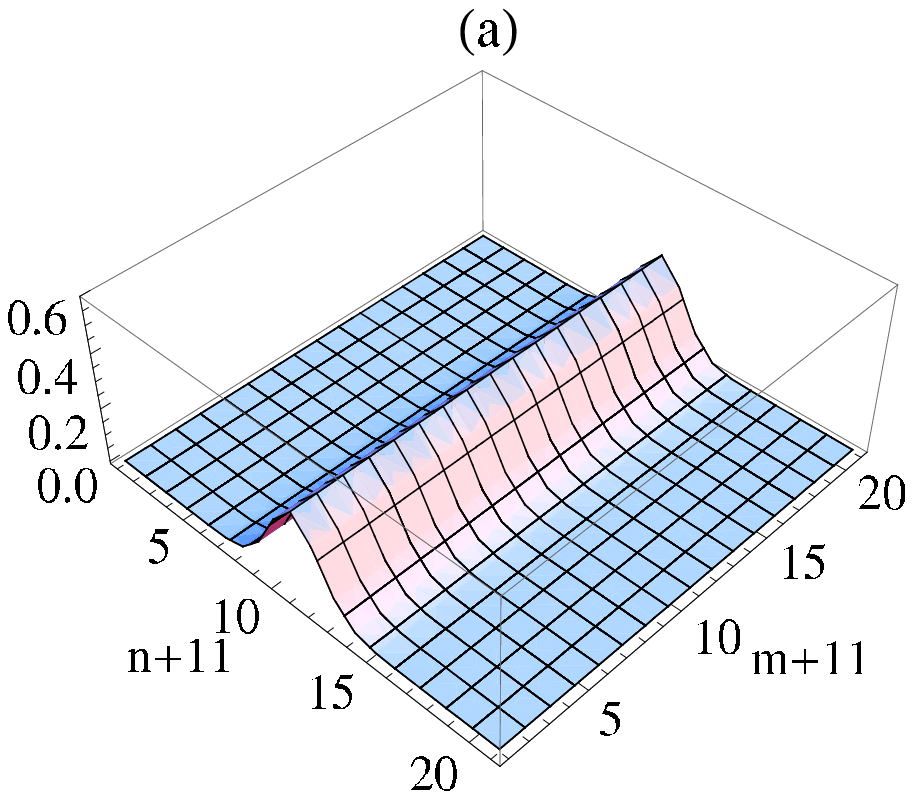}
\includegraphics[width=5.85cm]{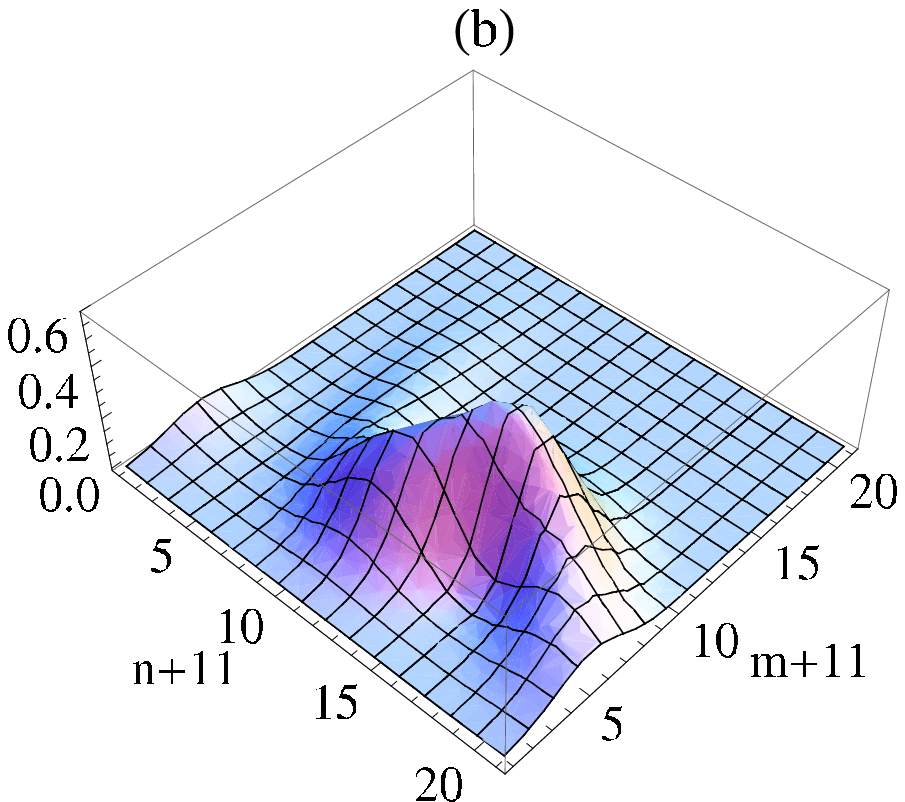}
\includegraphics[width=5.85cm]{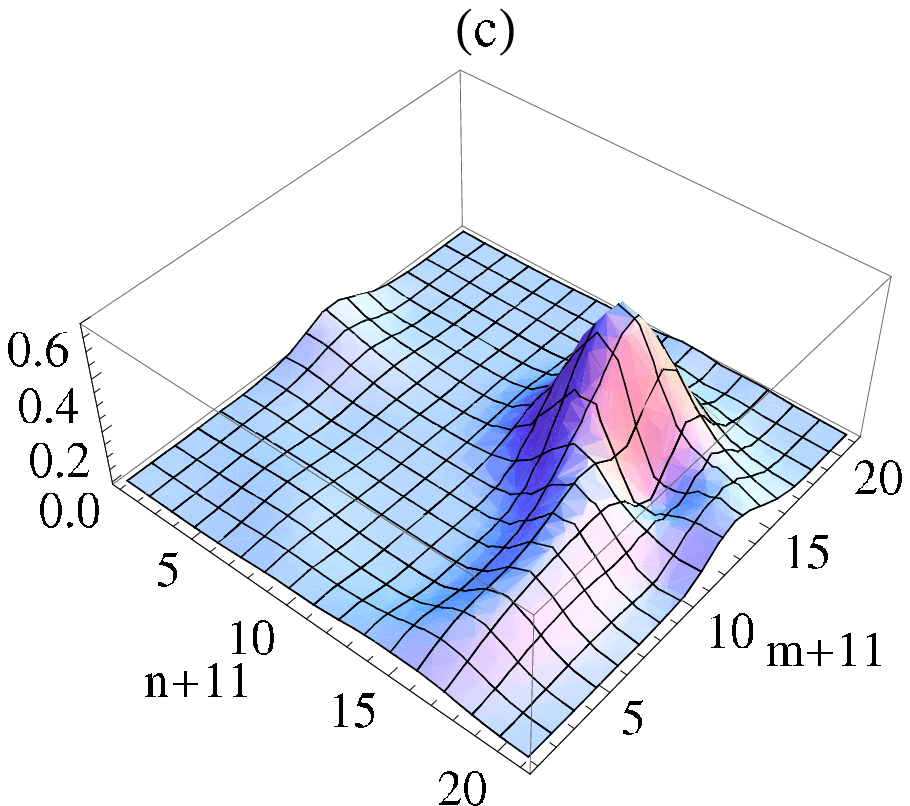}
\includegraphics[width=5.85cm]{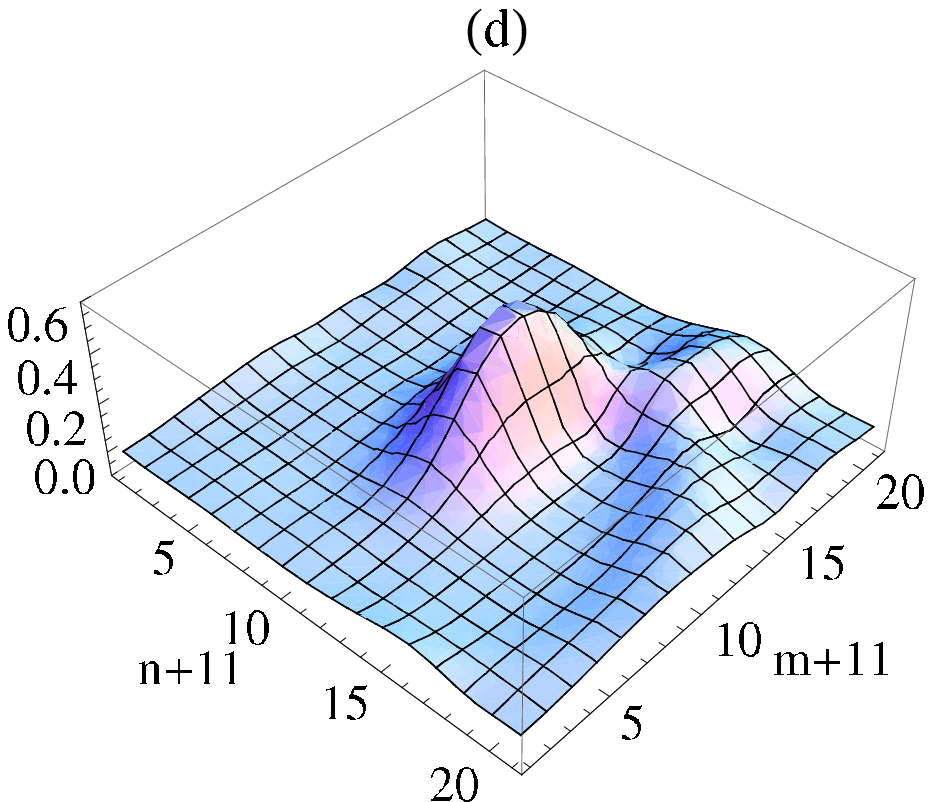}

\caption{Time evolved Husimi function under the same conditions indicated in Fig. 1.}

\end{figure}

In general, if the Wigner function evolves in time not clearly showing the
behavior of the physical system, the Husimi distribution, on the other hand, can
afford for a better visualization of how the state evolves over the discrete
phase space. In particular, it is to be noted the smeared behavior that the 
Husimi distribution presents.  

Also we see that, after a transient time -- during which the initial distribution loses its
identity --  the peak of the distribution passes recurrently on the same regions
of the $\theta =\left[ 0,\pi\right]$ sector of the discrete phase space because of the
quantum correlations and the presence of the transverse magnetic field in the $xy$-plane.

\subsection{Symmetric combination}
As a second case we consider the symmetric combination, namely,%
\begin{equation*}
\mid \psi ^{s}\rangle =\frac{1}{\sqrt{2}}\sum_{k}\left(
C_{k}^{i}|u_{k}^{i}\rangle +C_{k}^{j}|u_{k}^{j}\rangle \right) ,
\end{equation*}%
where the labels $i$ an $j$ characterize the $Fe8$ cluster energy
eigenstates of which the combination is made of. 
Here we will be interested in the energy doublet constituted of the  ground state and its neighbor partner.
This choice is based on the fact that this doublet is the proposed candidate for showing the spin tunneling
properties when the $Fe8$ cluster is submitted to temperatures below the well established crossover
 temperature. Furthermore, it is also known that the energy gap between
them is widened when an external magnetic field -- parallel to the easy-axis
-- is applied on the cluster. In particular, this study is of great interest when
this external magnetic field, $H_{\parallel }$, is an integer multiple of $0.22$ T because of
the matching of energy levels as pointed out before.

For those states it is also simple to obtain the Wigner function by means of
Eq. (\ref{wigner1}), namely $\cal{W}^{s}\left( m,n\right) $, and
propagate it in time. In Fig. 3 we depict a sequence of
time evolved Wigner functions for $H_{\parallel }=0.11\ T$ introduced in order to induce a convenient widening of the energy gap, and $H_{\perp }=0.0\ T$. For this situation we have, through the diagonalization of the Hamiltonian, the reference values $E_g = -29.01745\; K$ and $E_1 = -26.06441\; K$ which are the ground and the first excited states respectively. The energy gap beetwen those states is $\Delta E_{ref} = 2.95304\; K$.

\begin{figure}[h!]
\includegraphics[width=6.0cm]{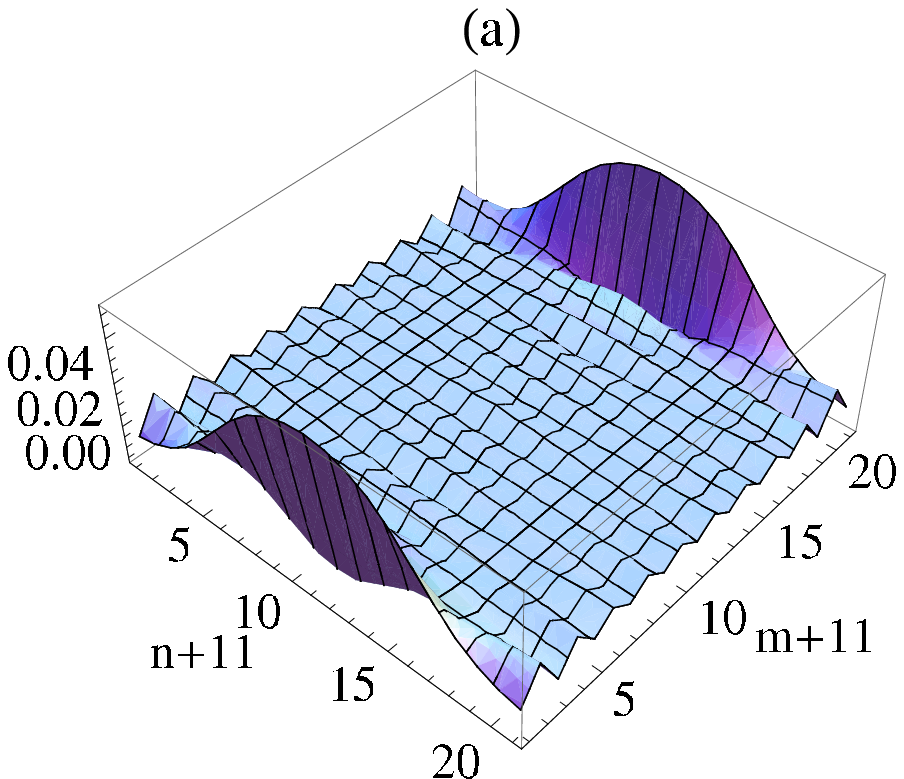}
\includegraphics[width=5.7cm]{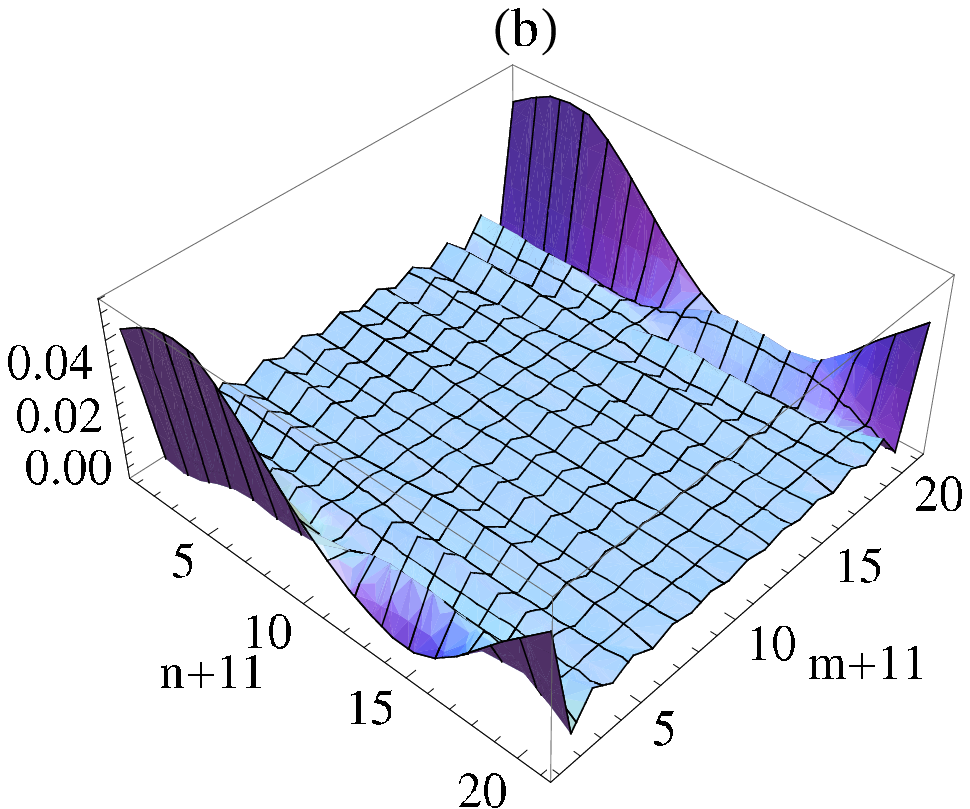}
\includegraphics[width=5.7cm]{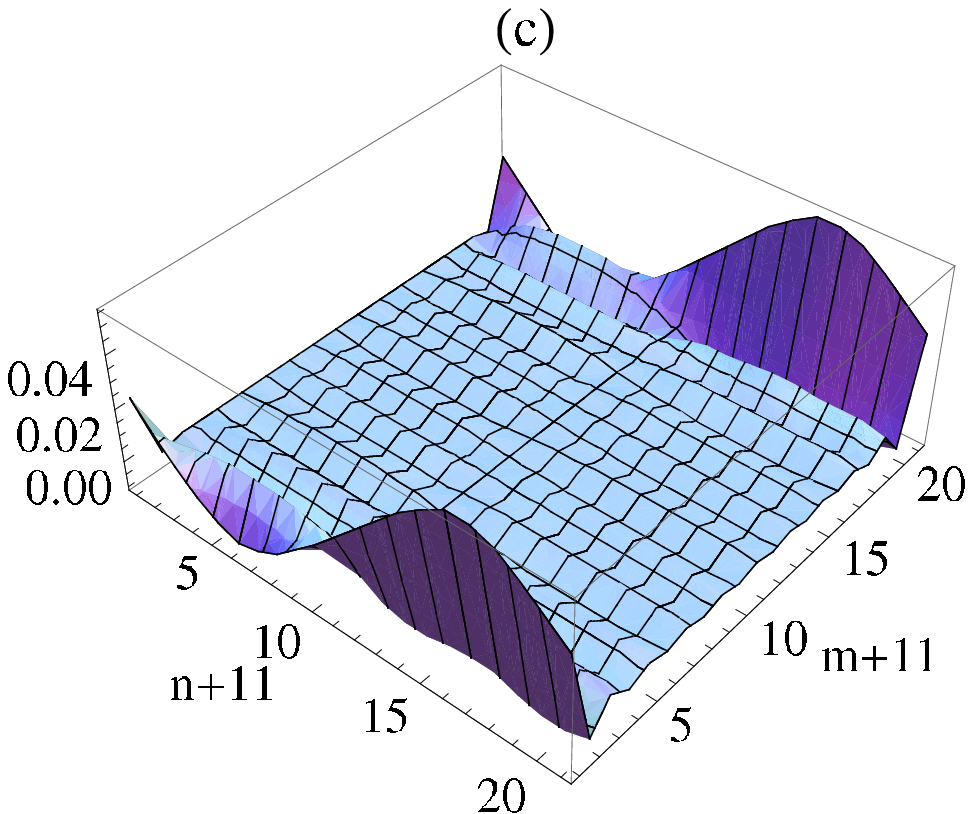}
\includegraphics[width=5.7cm]{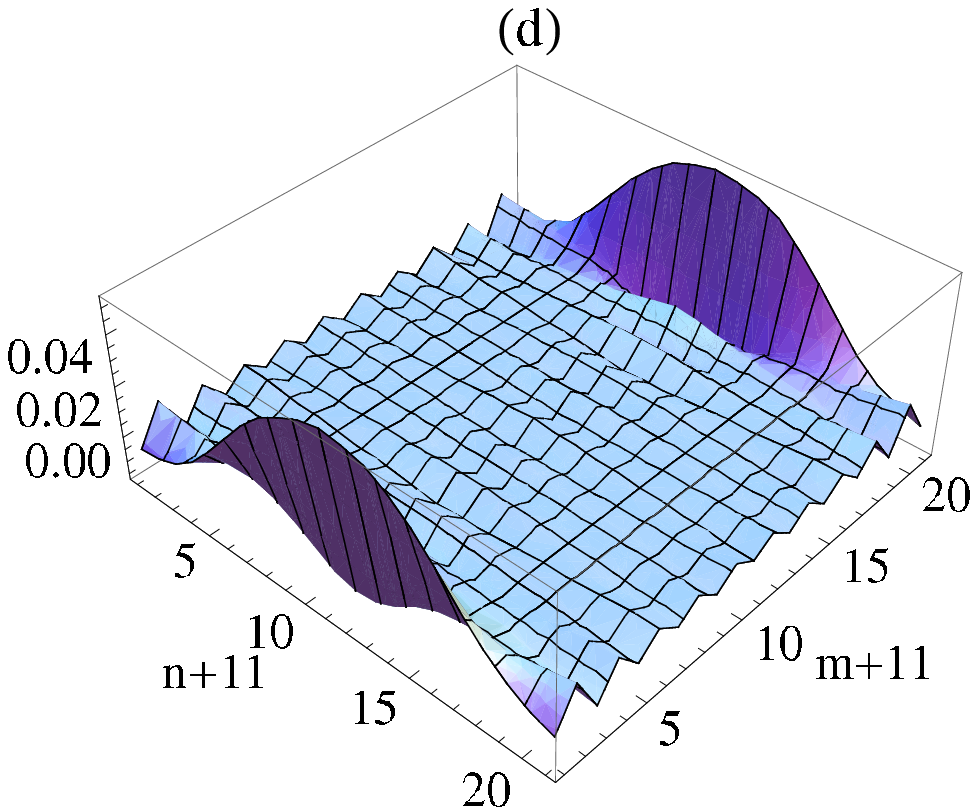}
\caption{Time evolution of the Wigner function generated by a symmetric combination of the lowest energy levels. The time step was taken as $\Delta$t = $0.05\times 2\pi\; K^{-1}$, and the external magnetic fields intensities were $H_{\parallel} = 0.11$ T and $H_{\perp} = 0.0$ T. Here the labels m and n also are the discrete angular momenta and angle variables. Figure (a) corresponds to the Wigner function at $\tau=0$. Figures (b), (c), and (d) depict the Wigner functions for $\tau = 0.75\times 2\pi\; K^{-1}$, $1.55\times 2\pi\; K^{-1}$ and $2.1\times 2\pi\; K^{-1}$ respectively.}
\end{figure}

First of all we observe that the staggered
behavior of the Wigner function, mainly visible in the central region of the
discrete phase space, is due to parity effects related to the angular
momentum variable. Moreover, what is important here is that the initial
configuration is totally recovered after $\tau \approx 2.15 \times 2\pi\; K^{-1}$, as can be
promptly recognized; this periodicity with time of the Wigner function 
indicates that we have a coherent oscillation that is associated with 
the magnetic moment quantum tunneling. 

If, instead of the Wigner function, we have the family of the corresponding Husimi distributions,
as shown in Fig. 4, we immediately see that the staggering was smeared.
Also, the oscillation can be clearly seen to occur along the angle sector of
the phase space such that the periodicity is obtained and agrees with that
observed with the Wigner function, in other words, the smearing process does not alter
the oscillatory character of the motion, so that the periodicity is the same.

\begin{figure}[h!]
\includegraphics[width=5.7cm]{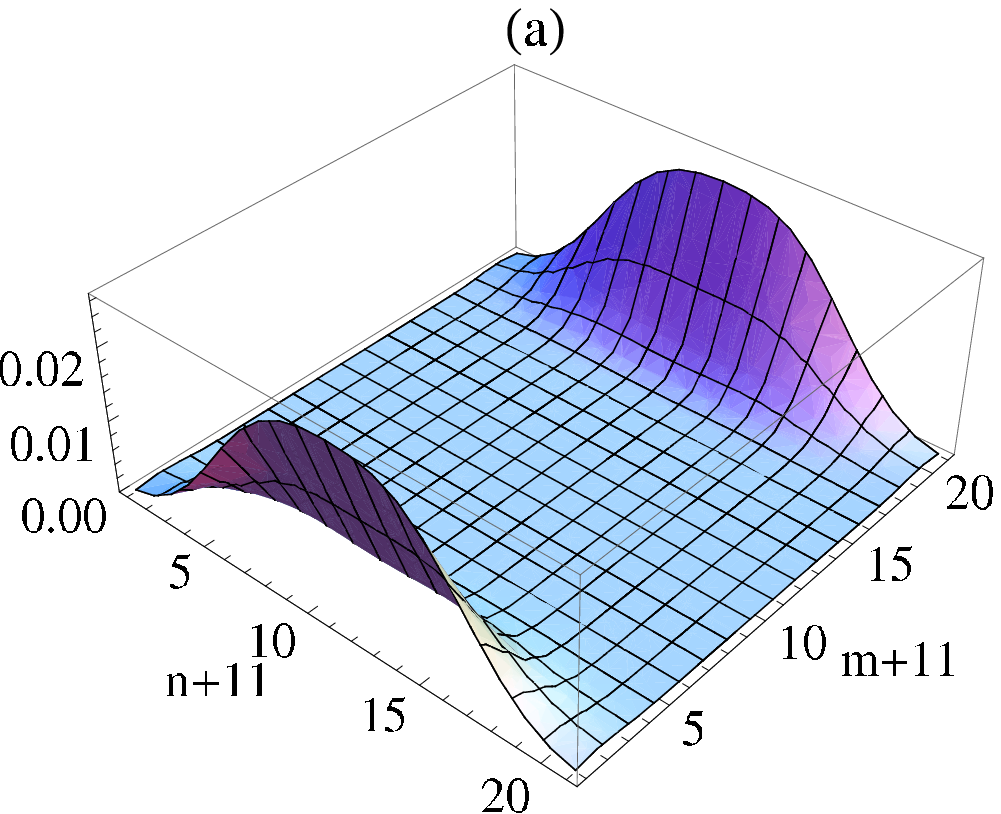}
\includegraphics[width=5.7cm]{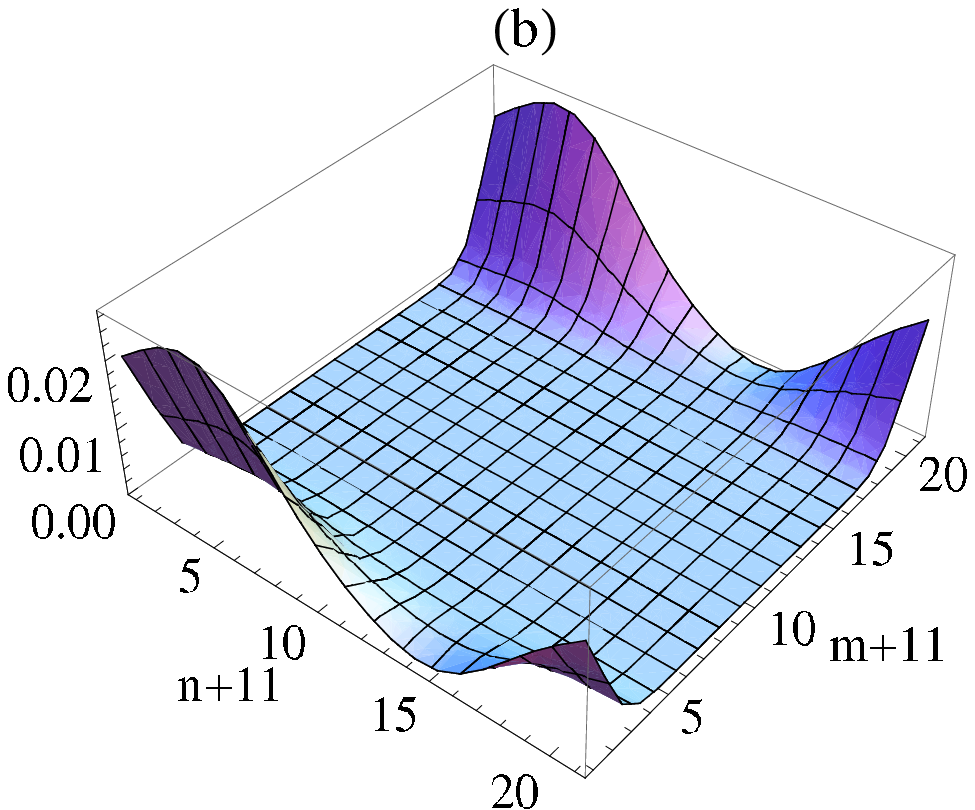}
\includegraphics[width=5.7cm]{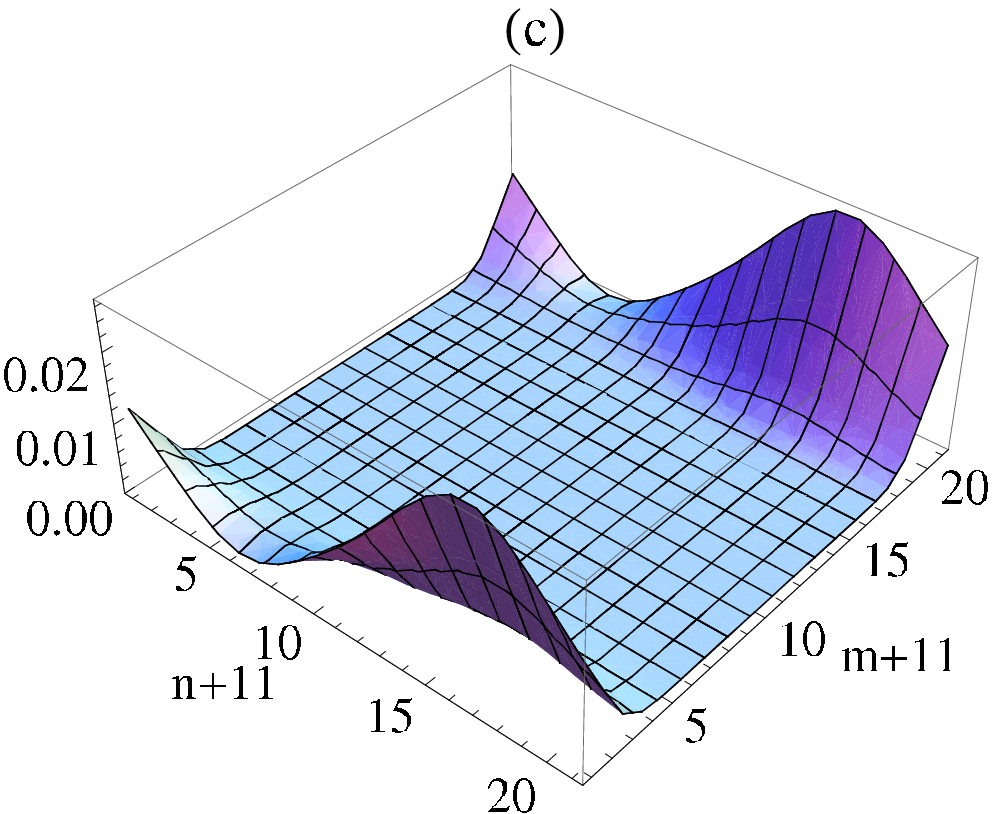}
\includegraphics[width=5.7cm]{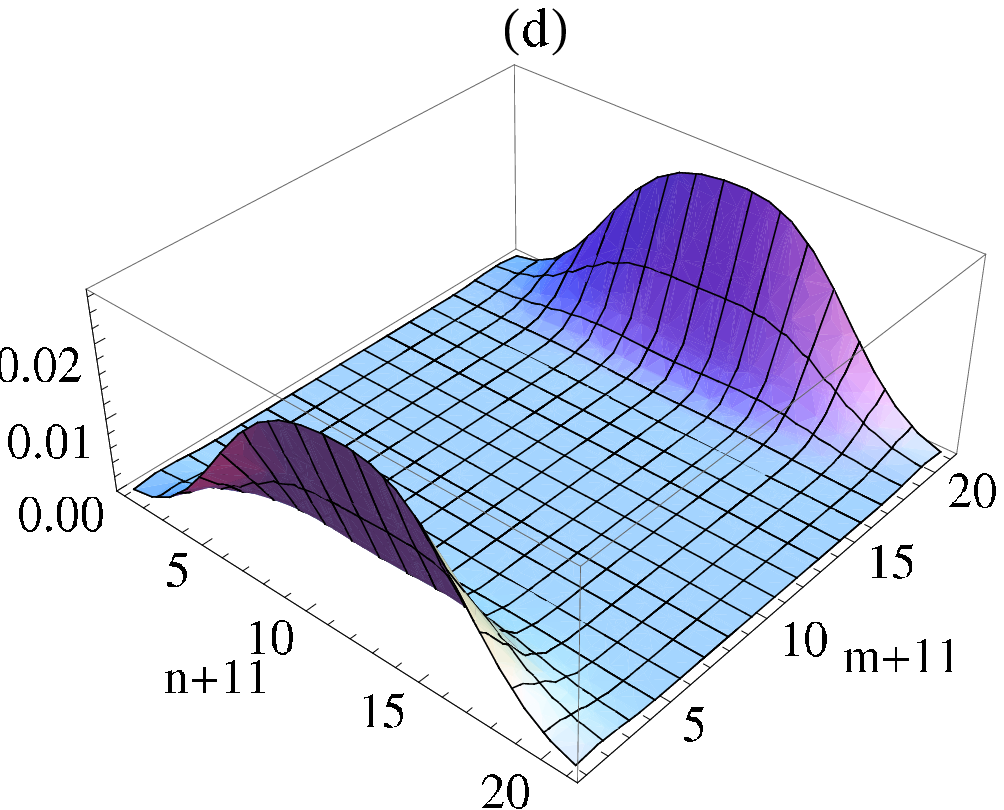}
\caption{Time evolved Husimi distribution under the same conditions indicated in Fig. 3.}
\end{figure}

Note that, the periodicity of the Wigner function
associated with that combination of states could also be obtained from the
time correlation function defined as
\begin{equation*}
P_{if}\left( \tau\right) =\sum_{m,n}\cal{W}^{i}\left( m,n;0\right) \cal{W}^{f}\left( m,n;\tau\right).
\end{equation*}%
A calculation performed with the same external magnetic field as before ($H_{\parallel
}=0.11\ T$ and $H_{\perp }=0.0\ T$) gives $P_{if}\left( \tau\right) $
as shown in Fig. 5.
\begin{figure}[h!]
\centering\includegraphics[width=6cm]{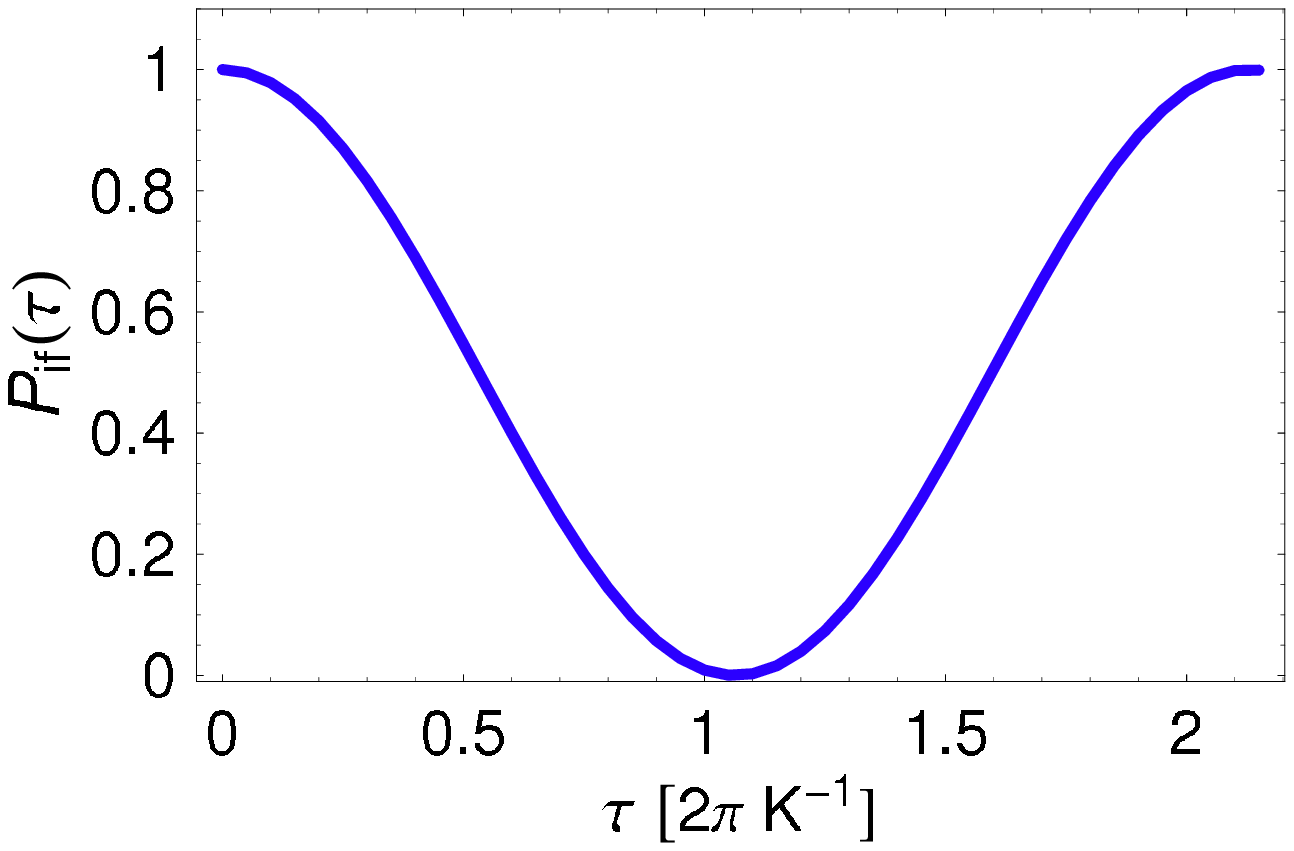}
\caption{Time correlation function calculated with time step of $\Delta$t = $0.05 \times 2\pi\; K^{-1}$. The external magnetic field intensities are $H_{\parallel} = 0.11$ T and $H_{\perp} = 0.0$ T.}
\end{figure}
Once again we obtain for the symmetric state a
periodicity in time of $\tau \approx 2.15 \times 2\pi\; K^{-1}$. Of course 
the oscillation period can be obtained from Fig. 4 as well from Fig. 5; both coincide.

Now, as an interesting result we see that we can associate this oscillation
period, $\tau \left( H_{\parallel }\right) $, with the energy gap between
the ground state and its neighbor partner -- which give the first energy
doublet -- by just using%
\begin{equation}
\Delta E\left( H_{\parallel }\right) =\frac{2\pi }{\tau \left( H_{\parallel
}\right) },  \label{gap}
\end{equation}%
where we can associate a frequency $\omega \left( H_{\parallel }\right) $
with the oscillation period, and $\Delta E\left( H_{\parallel }\right) $ is
the energy gap given as a function of the magnetic field intensity. It is 
important to emphasize that this relation is dependent on the magnetic
field intensity, and that here we want to consider only week magnetic field intensities
effects on the tunneling spin energy doublet. For the values of the above
considered external magnetic field and using $\tau
=2.15 \times 2\pi\; K^{-1}$, we obtain from Eq. (\ref{gap}) $\Delta E=2.92241\; K$, 
while the reference value is $\Delta E_{ref}= 2.95304\; K$, thus showing a
deviation of $1.04\%$, which indicates a good agreement.

Still along the same line we have just done, and in order to further clarify
the tunneling process ocurring in this case, we can also use an angle-based approach
developed in recent years \cite{dg,dges} to describe this same physical
situation. In that approach we obtain an effective Hamiltonian, written in
terms of a potential and an effective mass function, that can describe the
present situation and that can also complement the present discussion. To
highlight the main spin tunneling characteristics visualized in Figs. 4 
and 5 we plot the associated potential function, Fig. 6,
\beq
\nonumber V(\theta)&=&(D+E)J(J+1)\cos^2(\theta)
\\\nonumber &-& g\mu_B H_{||}\sqrt{J(J+1)}\cos(\theta)-EJ(J+1),
\eeq
\begin{figure}[h!]
\centering\includegraphics[width=6cm]{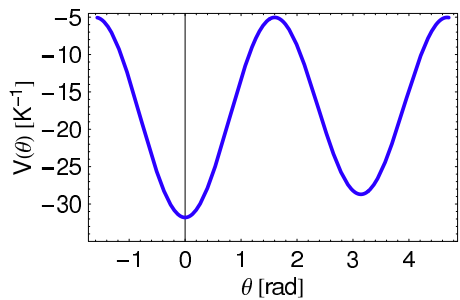}
\caption{Potential function of the Fe8 cluster obtained through an angle-based approach. The magnetic field intensities are $H_{\parallel} = 0.11$ T and $H_{\perp}=0.0$ T.}
\end{figure}
\newline where the Fe8 parameters are those already presented before. We then see
that the peaks of the wave functions which generate the Husimi distribution -- 
associated with the symmetric combination at $\tau=0$ -- occur at both deep minima
 at $\theta=0$ and $\theta=\pi$ of the potential. It is then
immediate to verify that the leakage of the Husimi distribution into angular
regions that would not be classically accessible indicates that tunneling is ocurring.
\subsection{Entropy Functionals}
As is well known, the Husimi distribution is an important tool to infer
the quantum information associated with the particular quantum state of the physical system it
represents. In this connection we can in particular write a Wehrl's type
entropy expression \cite{wherl} in terms of the Husimi distribution 
\begin{equation*}
S\left( \cal{H}\right) =-\f{1}{N}Tr\left[ \cal{H}\log \left[ \cal{H}\right] \right] .
\end{equation*}%
Now, since we can consider the Husimi distribution as a matrix over the
discrete phase space -- it assumes a positive value on every site of the
phase space --, we can write the entropy as \cite{march}
\begin{equation}
S\left( \cal{H};\tau\right) =-\f{1}{N}\sum_{m,n=-\ell}^{\ell}\cal{H}\left( m,n;\tau\right) 
\log \left( \cal{H}\left(m,n;\tau\right) \right) ,  \label{enthusimi}
\end{equation}%
where the sums are taken over the entire phase space, and it is immediate to
see that it is non-negative.

Taking that expression as our starting point, we can extract its inherent
physical interpretation by just choosing a particular quantum state of which
the essential features of the corresponding Husimi distribution have been
previously analysed. Moreover, since we are interested here in Husimi
distributions associated with simple pure states, we will focus our
attention on the time evolution of the entropy functional since it can give
us essential information about the time evolution of the particular state
all over the discrete phase space, in particular the indications of coherent
oscillations associated with spin tunneling.

As an emblematic case to be treated here, we consider the symmetric
combination of the lowest energy doublet of the Fe8 cluster submitted to an
external magnetic field parallel to the easy-axis. As has been seen before,
in the time evolution of this state a period of oscillation is recognized
that characterizes the energy gap in the spectrum. On the other hand,
this same fact can also be studied from the entropy point of view, what will
give us some indications of its interpretation \cite{marevaga}. Performing
the numerical calculations for $H_{\parallel }=0.11\ T$ and $H_{\perp }=0.0\
T$, we get the corresponding value of the entropy
at each step of the time propagation. The oscillatory character of the
entropy thus obtained is depicted in Fig. 7.

\begin{figure}[h!]
\centering\includegraphics[width=6cm]{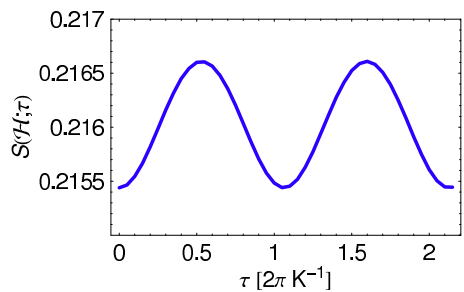}
\caption{Time dependence of the entropy function of the symmetric combination 
of the two lowest states using the same parameters as in Figs. 4 and 5.}
\end{figure}

Based on these results we verify that the Wehrl's type entropy in this case measures how the Husimi
distribution spreads over the angle and angular momentum sectors of the
discrete phase space. But we have to consider that the entropy is a minimum
whenever the Husimi distribution is maximally concentrated in a particular
region of the phase space, and is a maximum when the distribution spreads
over the phase space. Since the entropy does not distinguish negative or
positive sectors of the phase space, we have to consider for the period of
oscillation the time it takes for the system to start from a minimum in a
particular region of the phase space and to return to that same region.
Taking this into account, the energy gap can be calculated from the period thus obtained
and agrees again with the reference value within an error of $1.95\%$. 
Thus, by observing the oscillation of the entropy associated with the
symmetric -- or antisymmetric -- combination of an energy doublet we can
infer the gap energy information associated with the magnetic moment
tunneling.

Besides, we can also use the mutual correlation functional \cite{wherl,vedral} as well
\beq
\mathcal{I}(\mathcal{H};t) := S(\mathcal{J};t) + S(\Theta;t) 
- S(\mathcal{H};t) \geq 0 ,
\label{entmut}
\eeq
where $S(\mathcal{H};t)$ corresponds to time-dependent entropy defined in Eq. (\ref{enthusimi}), 
$S(\mathcal{J};t)$ and $S(\Theta;t)$ represent the marginal entropies which are related to the respective 
marginal distributions $\mathcal{J}(m;t)$ and $\Theta(n;t)$ \cite{marmauga,marevaga}, obtained by taking the trace of the corresponding Husimi distribution. With this functional we recover exactly the same behavior for the symmetric combination as that presented in Fig. 7, which allows us to infer possible correlations between the states and also calculate the energy gap as it was done through the Wehrl's type entropy.
\section{Conclusions}
Based on a formal approach aiming at the construction of a discrete phase
space picture of quantum mechanics related to finite-dimensional state
spaces developed before, and briefly reviewed here, we addressed the
problem of spin tunneling in the Fe8 magnetic cluster. Our formal approach
was shown to hold also in this particular case of  finite-dimensional quantum
system, and starting from the phenomenological Hamiltonian describing the
magnetic cluster with or without an external magnetic field we obtained the
discrete phase space representatives of the relevant elements which govern
the complete description of the system and its time evolution. The
Liouvillian associated with the Hamiltonian -- which propagates the density
operator in time in the discrete phase space -- was explicitly written. In
what concerns the representation of the state of the Fe8 magnetic cluster we
have shown how a discrete Wigner function, as well as a discrete Husimi
distribution, can be directly obtained.

The time evolution of those functions for the Fe8 magnetic cluster was studied in two different
cases: a) sharp angle state with an external magnetic field, and
b) symmetric combination of two energy eigenstates of the phenomenological
Hamiltonian; in particular, we chose the lowest energy doublet. In both
cases the Wigner function, as well as the Husimi distribution, were
calculated and relevant information concerning the dynamics of the spin was extracted from them. 
With respect to the time evolution of the
symmetric combination of states, it was also shown that, by analysing the
periodicity of its oscillatory behavior, a period could be found that
allowed us to calculate the energy gap associated with the tunneling doublet of states
from which we started.

Finally, we introduced a Wehrl's type entropy functional based on the discrete
Husimi distribution that allowed us to obtain a measure of the information
related to how the Husimi distribution is distributed over the discrete
phase space. Well concentrated distributions in phase space are associated
with small entropies, while the opposite produces greater entropies. The
coherent oscillations verified in the symmetric combination of the lowest energy
eigenstates thus reflect an oscillation in the associated time dependent
entropy. Therefore, it was also possible to extract the value of the energy
gap of the energy doublet through an analysis of the entropy. Furthermore, 
the values obtained via the mutual correlation functional lead us to the same physical 
interpretation as that obtained with the Wehrl´s type entropy functional.

The results presented here corroborate all the basic formal scheme proposed as well as the  
procedures adopted here to describe spin systems, in particular the Fe8 magnetic cluster,
in the corresponding discrete phase space.

\acknowledgments
The authors are thankful to M. A. Marchiolli for a careful reading of the manuscript and
for valuable suggestions. E. C. Silva has been supported 
by CAPES, Coordenaç\~{a}o de Aperfeiçoamento de Pessoal de N\'{i}vel Superior, Brazil.

\end{document}